\documentclass[doublecol]{epl2} 
\usepackage{graphicx}
\usepackage{rotating}
\usepackage{amsmath}
\usepackage{amsfonts}
\usepackage{amssymb}
\usepackage{enumerate}
\usepackage{longtable}
\usepackage{units}
\usepackage{hyperref}
\usepackage[T1]{fontenc}
\usepackage[latin1]{inputenc}
\usepackage{graphicx}
\usepackage{capt-of}
\usepackage{bbm}
\usepackage{dcolumn}% Align table columns on decimal point
\usepackage{bm}
\usepackage{color}

\begin{document}
\newcommand{\cred}{\color{red}}

\title{A generalized perspective on non-perturbative linked cluster expansions}

\author{K. C\"oster\inst{1} \and S. Clever\inst{1} \and F. Herbst\inst{1} \and S. Capponi\inst{2} \and K. P. Schmidt\inst{1}}
\shortauthor{K. C\"oster, S. Clever, F. Herbst, S. Capponi,  and K. P. Schmidt}

\institute{
\inst{1} Lehrstuhl f\"ur Theoretische Physik I, Otto-Hahn-Stra\ss e 4, TU Dortmund, D-44221 Dortmund, Germany\\
\inst{2} Laboratoire de Physique Th\'eorique, CNRS UMR 5152, Universit\'e Paul Sabatier, F-31062 Toulouse, France
}

\date{\rm\today}


\abstract{
We identify a fundamental challenge for any non-perturbative approach based on finite clusters resulting from the reduced symmetry on graphs, most importantly the breaking of translational symmetry, when targeting the properties of excited states. This can be traced back to the appearance of \emph{intruder states} in the low-energy spectrum, which represent a major obstacle in quasi degenerate perturbation theory. Here a generalized notion of cluster additivity is introduced, which is used to formulate an optimized scheme of graph-based continuous unitary transformations allowing to solve and to physically understand this major issue. Most remarkably, our improved scheme demands to go beyond the paradigm of using the exact eigenvectors on graphs. 
}

\maketitle

%
%
%%%%%%%%%%%%%%%%%%%%%%%
%%%%%%%%%%%%%%%%%%%%%%%
\emph{Introduction ---}
%%%%%%%%%%%%%%%%%%%%%%%
%%%%%%%%%%%%%%%%%%%%%%%
%
%
The collective behaviour of quantum matter is one of the most fascinating topics of modern physics. Understanding it is crucial, as it holds the key to a variety of correlated many-body states - realised for example in spin liquids or superconductors. In particular, it is decisive to gain a systematic understanding of collective phenomena to identify fundamentally new behaviour. One attractive route is to study the physics of strongly interacting quantum lattice models, which very often demands the use of efficient numerical tools in order to gain quantitative insights. 

Besides exact diagonalizations, quantum Monte Carlo simulations, or variational tensor network calculations, so-called linked cluster expansions (LCEs) became, over the last decades, a standard tool to study quantum many-body systems \cite{Hamer_Book}. Based on the linked-cluster theorem, high-order series expansions of various physical quantities can be evaluated directly in the thermodynamic limit by performing calculations on finite systems. The latter include zero-temperature properties like the ground-state energy \cite{Maryland81,Irving84}, order parameters, or entanglement entropies \cite{Singh12} as well as high-temperature expansions giving access to thermodynamic quantities \cite{He90}. Interestingly, it took until 1996 to set up similar expansions for the physical properties of elementary excitations like one-particle dispersions \cite{Gelfand96}, two-particle interactions \cite{Trebst00,Zheng01,Knetter03_1} or dynamical correlation functions \cite{Knetter01,Schmidt01}, i.e.~physical properties which are of direct importance for the interpretation of inelastic neutron or inelastic light scattering experiments.     

The usefulness of high-order series expansions is limited due to its perturbative nature. It is therefore desirable to reflect about non-perturbative linked cluster expansions (NLCEs) \cite{Rigol06,Rigol07_1,Rigol07_2,Rigol11,Tang13}.  The essential idea behind all NLCEs is a non-perturbative treatment of graphs, achieved via an exact (block) diagonalization, yielding results in the thermodynamic limit after an appropriate embedding procedure. Indeed, many exciting developments have been achieved in this direction recently, e.g.~the derivation of effective low-energy spin models \cite{Yang11,Yang12, Ixert14}, the calculation of entanglement entropies \cite{Kallin13} or the extension to time-dependent quantities out of equilibrium \cite{Rigol14,Rigol14_2}. Hence, NLCEs are a promising tool for the investigation of quantum lattice models with a vast range of applications without finite-size effects.

All LCEs share at their core that the physical system on clusters has a reduced symmetry compared to the infinite system, e.g.~the translational symmetry is broken by construction. For the perturbative LCEs, the full symmetry is nevertheless restored after embedding, and exactly the same fluctuations present in the thermodynamic limit are taken into account. However, as we demonstrate in this letter, this inherent symmetry reduction represents generically a fundamental challenge for any non-perturbative approach based on finite clusters when calculating excitation energies of elementary excitations. Here we introduce a generalized notion of cluster additivity which allows to solve this challenge by an adapted version of graph-based continuous unitary transformations (gCUTs) \cite{Yang11}. Fascinatingly, this generalization requires not to use the exact eigenvectors on graphs, revealing once more the non-trivial connection between finite systems and the thermodynamic limit in quantum many-body systems.

%
%
%%%%%%%%%%%%%%%%%%%%%%%
%%%%%%%%%%%%%%%%%%%%%%%
\emph{Set up ---}
%%%%%%%%%%%%%%%%%%%%%%%
%%%%%%%%%%%%%%%%%%%%%%%
%
%
We consider a generic quantum lattice Hamiltonian $\cal{H}$ at zero temperature. By decomposing the original lattice into a superlattice of supersites, one can always rewrite \emph{exactly} ${\cal H}$ as ${\cal H}={\cal H}_0+\lambda {\cal V}$ \cite{Huerga13}. Here a supersite might be a spin, two linked spins like a dimer, or any other finite set of linked sites which can be easily diagonalized. In the following we focus on elementary excitations of such quantum lattice models.

The part \mbox{$\mathcal{H}_0=E_0+\sum_{i,\alpha}a_0^\alpha \hat{f}^\dagger_{i,\alpha}\hat{f}^{\phantom{\dagger}}_{i,\alpha}$} is diagonal in supersites $i$ of the lattice containing the local quantum degrees of freedom which interact via short-range operators of $\mathcal{V}$ building the bonds of the lattice. $E_0$ denotes a constant and the sum runs over all supersites $i$ and all local excitations $\alpha$. Generically, $\mathcal{H}$ is expressed in normal-ordered form with respect to $\mathcal{H}_0$ as $\mathcal{H}=\mathcal{H}_{\rm c}+\mathcal{H}_{\rm nc}$ using annihilation (creation) operators $\hat{f}_{i,\alpha}^{(\dagger)}$ 
\begin{eqnarray}
 \mathcal{H}_{\rm c} &=& E_0(\lambda) + \hspace*{-2mm}\sum_{i,\delta,\alpha,\beta} a_\delta^{\alpha\beta}(\lambda)\, \hat{f}^\dagger_{i+\delta,\alpha}\hat{f}^{\phantom{\dagger}}_{i,\beta} + {\rm h.c.}+\ldots\\
 \mathcal{H}_{\rm nc} &=& \sum_{i,\delta,\alpha,\beta} \Gamma_\delta^{\alpha\beta}(\lambda)\, \hat{f}^\dagger_{i+\delta,\alpha}\hat{f}^{\dagger}_{i,\beta} + {\rm h.c.} + \ldots\, ,
\end{eqnarray}
where dots refer to other particle-conserving (non-conserving) terms in $\mathcal{H}_{\rm c}$ ($\mathcal{H}_{\rm nc}$). The goal is then to derive a renormalized particle-conserving Hamiltonian $\tilde{\mathcal{H}}_{\rm c}$ accounting for the influence of $\mathcal{H}_{\rm nc}$ quantitatively. This represents a substantial simplification, since fundamental quantities like a one-particle dispersion can be determined straightforwardly. The derivation of $\tilde{\mathcal{H}}_{\rm c}$ is well defined as long as no quantum phase transition occurs as a function of $\lambda$.

%
%
%%%%%%%%%%%%%%%%%%%%%%%
%%%%%%%%%%%%%%%%%%%%%%%
\emph{LCEs and cluster additivity ---}
%%%%%%%%%%%%%%%%%%%%%%%
%%%%%%%%%%%%%%%%%%%%%%%
%
%
The general concept behind any LCE is to decompose physical quantities in the thermodynamic limit into a sum of reduced contributions from finite linked clusters. We define a cluster of the infinite system as a finite subset of supersites and their linking bonds. The reduced contribution of a cluster is then obtained by subtracting reduced contributions of all subclusters to avoid double counting. Consequently, a reduced contribution corresponds to the fluctuations which are specific to a given cluster. 
%|||||||||||||||||||||||||||||||||||||||||||||||||||||||||||||||||||||||||||||||||||||||||||||||||||||||||||||||
\begin{figure}
\begin{center}
% NOTICE: lighter JPEG version below. Use EPS one for final version
\includegraphics*[width=0.9\columnwidth]{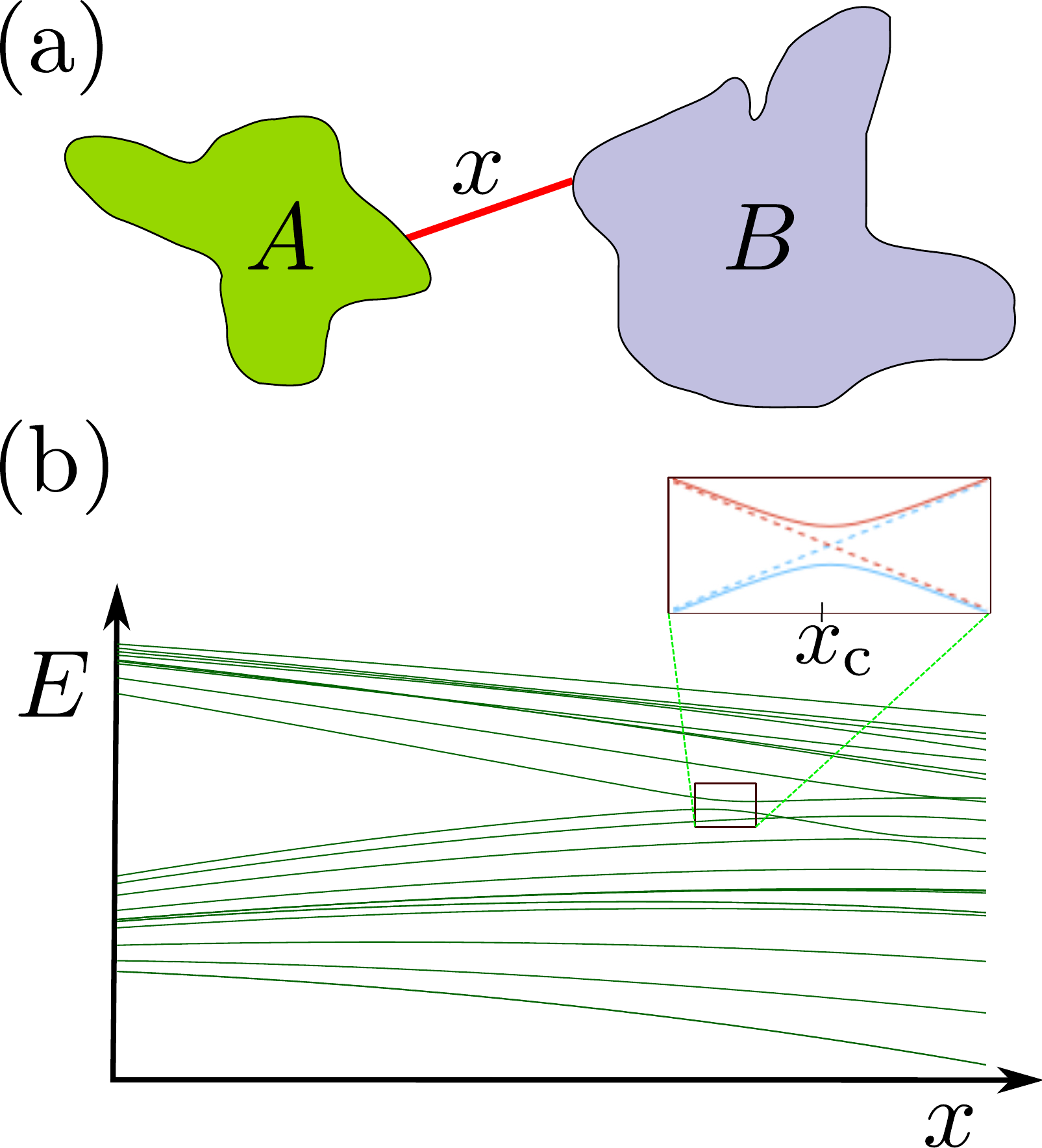}
\end{center}
\caption{{\it (Color online)} (a) Sketch of two clusters $A$ and $B$ which are linked by a single bond of strength $x$. (b) Sketch of the eigenvalues $E$ as a function of $x$ for the joined system $A+B$ linked by a single bond. ({\it Inset}) Zoom on the first anti-level crossing. Dashed lines represent the appropriate evolution of eigenvalues respecting the generalized notion of cluster additivity.} 
\label{fig:cluster_additivity}
\end{figure}
%|||||||||||||||||||||||||||||||||||||||||||||||||||||||||||||||||||||||||||||||||||||||||||||||||||||||||||||||
The latter is based on the so-called cluster additivity which is defined as follows. Let us call two clusters $A$ and $B$ disconnected, if they do not have any sites in common and there is no bond linking sites from cluster $A$ and $B$ (see Fig.~\ref{fig:cluster_additivity}a for $x=0$). For a disconnected cluster $C=A\cup B$ any quantity $\mathcal{M}^C$ is called \textit{cluster additive} if it can be expressed as
\begin{align}
\mathcal{M}^C:=\mathcal{M}^A \otimes \mathbbm{1}^B+\mathbbm{1}^A\otimes \mathcal{M}^B\,,
\end{align}
so that $\mathcal{M}^C$ splits into a part associated with the Hilbert space $\mathfrak{H}^A$ ($\mathfrak{H}^B$) of subcluster $A$ ($B$). The Hamiltonian $\mathcal{H}^C$ as well as $\tilde{\mathcal{H}}^C_{\rm c}$ are cluster additive, which implies a proper LCE for the ground-state energy $\tilde{E}$ and for the one-particle hopping amplitudes $\tilde{a}_\delta$. Overall, the cluster additivity allows therefore to unambiguously identify the reduced contributions of clusters and thus to consistently define the embedding into the infinite system.

%
%
%%%%%%%%%%%%%%%%%%%%%%%
%%%%%%%%%%%%%%%%%%%%%%%
\emph{Generalized cluster additivity ---}
%%%%%%%%%%%%%%%%%%%%%%%
%%%%%%%%%%%%%%%%%%%%%%%
%
%
Cluster additivity is sufficient to perform an NLCE if the reduced symmetry of the physical system on graphs before embedding into the infinite system does not matter. But this is not always the case. Let us consider a physical system where the one-particle mode is stable, i.e.~the one-particle dispersion is below any many-particle continuum for any momentum, and the maximum of the one-particle dispersion is larger than the minimum of many-particle energies. Consequently, it is the translational symmetry of the system which protects the one-particle mode from decaying. But this protection is a priori not at work on open clusters and one therefore expects an artificial entanglement between certain states. In such a situation, one has to generalize the notion of cluster additivity as we introduce in the following.

For clarification we consider a cluster $C$ consisting of two subclusters $A$ and $B$ which are linked by only one link of coupling strength $x$ as illustrated in Fig.~\ref{fig:cluster_additivity}a, and one aims at evaluating the reduced contribution for $x=\lambda$. Any quantity $\mathcal{M}^C$ like the Hamiltonian $\tilde{H}^C_{\rm c}$ takes then the form
\begin{align}
 \mathcal{M}^C=\mathcal{M}^A \otimes \mathbbm{1}^B+\mathbbm{1}^A\otimes \mathcal{M}^B+\mathcal{M}_{x},
\end{align}
where $\mathcal{M}_{x}$ gives rise to the reduced contribution of cluster $C$. One prototypical evolution of the eigenenergies of $\tilde{H}^C_{\rm c}$ are displayed in Fig.~\ref{fig:cluster_additivity}b. One observes two regimes: For small $x$, the change of eigenvalues and eigenvectors is smooth while, for larger $x$, there are characteristic anti-level crossings starting from $x_{\rm c}$ where eigenenergies and eigenvectors change drastically. If the corresponding fluctuations are not present in the thermodynamic limit, e.g.~a pseudo decay of a quasi particle only present on finite clusters as described above, these anti-level crossings can affect $\mathcal{M}^C$ severely in an unphysical fashion leading to the breakdown of NLCEs for $\lambda\gtrsim x_{\rm c}$.

One therefore has to generalize the notion of cluster additivity by demanding that any physical quantity $\mathcal{M}^C$ like $\tilde{H}^C_{\rm c}$ is sufficiently smooth as a function of $x$ for \mbox{$0\leq x\leq\lambda$} with respect to the thermodynamic limit. For a single anti-level crossing at $x_{\rm c}$ this can be naturally discussed by focusing on the two involved energy levels as sketched in the zoom of Fig.~\ref{fig:cluster_additivity}b. Denoting by $|i\rangle$ and $|j\rangle$ the two eigenvectors for $x<x_{\rm c}$, the eigenvectors at (or close to) $x_{\rm c}$ are entangled superpositions of $|i\rangle$ and $|j\rangle$. Artificial anti-level crossings should now be replaced by true level crossings as indicated by the dashed lines in Fig.~\ref{fig:cluster_additivity}b, since the two involved levels have different quantum numbers in the thermodynamic limit implying a smooth behaviour of all quantities. In other words, one has to systematically avoid unphysical entanglement and associated anti-level crossings. As a direct consequence, one cannot use the exact eigenvectors as it is done so far in all implementations of \mbox{NLCEs.}

%
%
%%%%%%%%%%%%%%%%%%%%%%%
%%%%%%%%%%%%%%%%%%%%%%%
\emph{Solving the challenge ---} 
%%%%%%%%%%%%%%%%%%%%%%%
%%%%%%%%%%%%%%%%%%%%%%%
%
%
A systematic disentanglement of specific levels involved in artificial anti-level crossings is
a priori a very complicated task and it is by far not obvious that a general solution exists. Nevertheless, in the following we formulate an optimized version of gCUTs taking fully into account the generalized notion of cluster additivity which therefore solves the fundamental challenge stated above.

The gCUT \cite{Yang11} is a non-perturbative variant of continuous unitary transformations (CUTs) \cite{Wegner94,Glazek93,Glazek94} where the full CUT in the thermodynamic limit is rephrased as an NLCE by summing up reduced contributions of graphs (for other variants see also Refs.~\cite{Knetter00,Krull12}). In practice, only a finite set of graphs can be treated numerically which sets a characteristic length scale $\mathcal{L}$ of quantum fluctuations captured. Consequently, if the physical system has a finite correlation length $\xi$ the gCUT converges as long as $\xi\sim\mathcal{L}$. Therefore, at quantum critical points with $\xi\rightarrow\infty$, one relies on appropriate scalings in $\mathcal{L}$. 

A CUT maps the initial Hamiltonian $\mathcal{H}$ by \mbox{$\mathcal{H}(\ell)=U^\dagger(\ell)\mathcal{H}U(\ell)$} unitarily using the continuous flow parameter $\ell$. The goal is to derive an effective, in our case quasi-particle conserving, Hamiltonian \mbox{$\tilde{\mathcal{H}}_{\rm c}\equiv\mathcal{H}(\ell=\infty)$} via the flow equation \mbox{$\partial_\ell \mathcal{H}(\ell)=[\eta(l),\mathcal{H}(\ell)]$} where \mbox{$\eta(\ell)$} is the antihermitian generator of $U(\ell)$. In gCUTs the flow equation is solved numerically on each graph, which is done exactly due to the finite Hilbert space dimension \cite{Yang11}. The graph-dependent effective Hamiltonian matrix $\tilde{\mathcal{H}}^C_{\rm c}$ is then embedded into the infinite system yielding the effective Hamiltonian $\tilde{\mathcal{H}}_{\rm c}$ in second quantization. 

Here we use on each cluster $C$ the quasi-particle (QP) counting operator $\hat{Q}^C(\ell)=\sum_i\hat{n}_i(\ell)$ with $\hat{n}_i=f^{\dagger}_i f^{\phantom{\dagger}}_i$ (restricting to a single $\alpha$ in $\mathcal{H}_0$), and define the QP-generator \mbox{$\eta^C_{i,j}=\text{sgn}(q_i-q_j) h^C_{i,j}$} with $\hat{Q}^C|i\rangle=q_i|i\rangle$ \cite{Knetter00,footnote_alpha}. For this generator the effective Hamiltonian matrix $\tilde{\mathcal{H}}^C_{\rm c}$ is block-diagonal in the numbers of QPs. We focus on the 1QP sector, and we set $q=0$ for 0QP, $q=1$ for 1QP, and $q=2$ for all other channels. The truncation of the cluster basis is crucial to treat large graphs. Without any form of truncation, the method is limited to smaller graph sizes, because a matrix of size $D_{\mathcal{H}}\times D_{\mathcal{H}}$ must be stored, where $D_{\mathcal{H}}$ denotes the Hilbert space dimension of the cluster under consideration. Here we use a block-Lanczos algorithm along the lines described in Ref.~\cite{Ixert14} which allows to treat graphs of length $L=12$ (size $N=22$) for the two-leg Heisenberg ladder (transverse-field Ising model on the square lattice).

For $q=1$, let us consider the $L$ states with $q=1$ of $\mathcal{H}_{\rm c}^{C}$ on cluster $C$. These states constitute the initial block in our block-Lanczos algorithm. Now one acts successively with the full Hamiltonian $\mathcal{H}^{C}$ generating a new set of states which, after orthonormalization with respect to all previous states, build the next block of our optimal cluster basis. This procedure is performed until the difference in the $q=1$ block of $\tilde{\mathcal{H}}_{\rm c}^{C}$ is negligible. Note that the states of the initial block remain unchanged during the generation of the cluster basis. For the calculation of the ground-state energy ($q=0$), our scheme simply reduces to a standard Lanczos algorithm for the lowest eigenenergy of a given cluster $C$.

The gCUT is expected to converge if no artificial anti-level crossings are present in the spectrum of the graphs considered. This is for example the case for the single-valued 0QP sector containing $\tilde{E}_0$. For this (scalar) quantity, gCUTs is exactly equivalent to NLCEs using exact diagonalization.   

The situation is more complex in the 1QP sector when artificial anti-level crossings are present between 1QP and $n$QP ($n>1$), since the QP-generator follows {\it always} the lower branch of an anti-level crossing on each cluster. This implies a breakdown of the gCUT for \mbox{$\lambda\gtrsim\lambda_{\rm c}$}, when $\lambda_{\rm c}$ denotes the location of the first artificial anti-level crossing. Such avoided level crossings are known to represent a major obstacle in quasi degenerate perturbation theory, since either one always picks the lowest state and gets non-reliable results, or one chooses instead the diabatic procedure which results in discontinuous properties.~\cite{Malrieu1985,Evangelisti1987}

Our aim is therefore to perform a CUT which does not fully separate the 1QP sector from the rest of the states, so that elements $h_{ij}$, giving rise to artificial anti-level crossings, stay finite after the CUT \cite{footnote1}. Interestingly, it is the additional "flow dimension" $\ell$ which gives the freedom to do so.

\begin{figure}
\begin{center}
% NOTICE: lighter JPEG version below. Use EPS one for final version
\includegraphics*[width=\columnwidth]{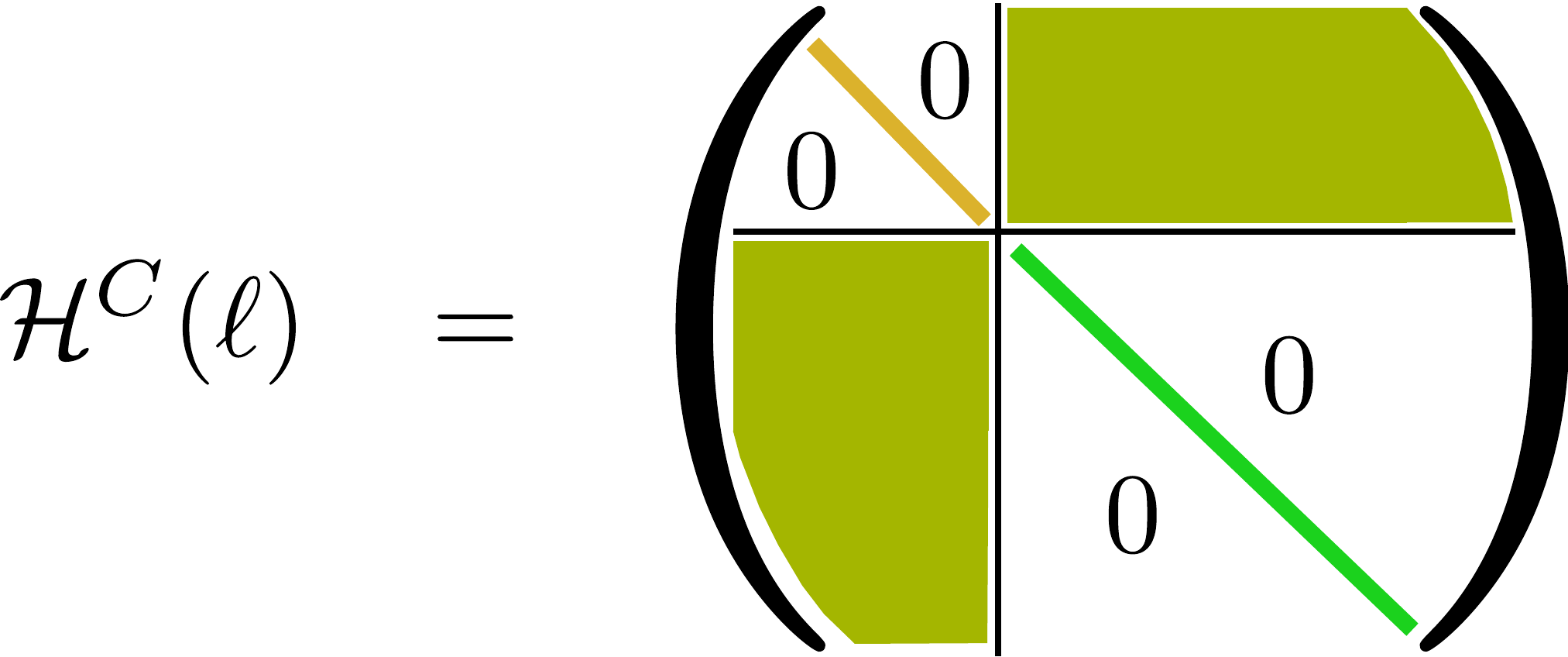}
\end{center}
\caption{{\it (Color online)} Sketch of $\mathcal{H}^{C}(\ell)$ in the diagonal basis of sub-blocks. The upper left (lower right) diagonal block contains the sub-block eigenenergies $\epsilon_i^{C}$ of the sub-block eigenstates $|1\rangle$ ($|n\rangle$) with $q=1$ ($q=2$). The matrix elements $h_{1n}$ between such sub-block eigenstates constitute the filled off-diagonal blocks.} 
\label{fig:diagonal_basis}
\end{figure}

To this end let us consider a one-quasi-particle ($q=1$) eigenstate $|1\rangle$ and an $n$-quasi-particle ($q=2$) eigenstate $|n\rangle$ during the flow in the diagonal basis of sub-blocks (see Fig.~\ref{fig:diagonal_basis}), i.e.~any QP-sector is diagonalized disregarding the finite matrix elements between different sectors. This is the optimal basis to investigate anti-level crossings, since the interaction between sub-block eigenstates $|1\rangle$ and $|n\rangle$ with different number of QPs is contained in the single matrix element $h_{1n}(\ell)$. Physically, one disentangles the states $|1\rangle$ and $|n\rangle$ involved in the anti-level crossings as sketched in the inset of Fig.~\ref{fig:cluster_additivity}b. This is achieved by adapting the generator during the flow as follows. Typically, states $|1\rangle$ and $|n\rangle$ are almost unentangled in the window $0\leq\ell\leq\ell_{1n}^{\rm c}$, since the purely non-perturbative artificial entanglement between both states builds up significantly only for rather large values $\ell\gtrsim\ell_{1n}^{\rm c}$. Consequently, one sets $\eta_{1n}=0$ at $\ell_{1n}^{\rm c}$ for each artificial anti-level crossing. The corresponding element $h_{1n}(\ell=\infty)$ is then finite representing the artificial interaction due to the reduced graph symmetry. Finally, the $\ell_{1n}^{\rm c}$ are located using reduced weights \mbox{$W_{\nu}^C$} defined for any sub-block eigenstate $|\nu\rangle$ which are natural quantities in contractor renormalization group (CORE) approaches \cite{Morningstar96,Morningstar96b,Capponi04}. The reduced weights \mbox{$W_{\bar{\nu}}^C= \langle \bar{\nu}| \bar{\nu}\rangle $} are defined for any state as
\begin{equation}
\label{eq:weights}
| \bar{\nu}\rangle = \left(1-\sum_{\bar{\mu}} \frac{|\bar{\mu}\rangle \langle \bar{\mu}|}{W_{\bar{\mu}}}\right)   \frac{P|\tilde{\nu}\rangle}{\langle \tilde{\nu}|P|\tilde{\nu}\rangle}\quad ,
\end{equation}
where $|\tilde{\nu}\rangle$ represents $|\nu\rangle$ in the original basis at $\ell=0$ and $P$ projects to the one-particle states of $\mathcal{H}_{\rm c}(\ell=0)$. For 1QP ($n$QP), the sum runs over all 1QP states $|\bar{\mu}\rangle$ having a (sufficiently) smaller sub-block energy.

The physical logic behind the $W_{\bar{\nu}}^C\in[0,1]$ becomes apparent for $\ell=\infty$ as a function of $\lambda$. If $\lambda\ll \lambda_{\rm c}$, then all $W_{\bar{\nu}}^C$ of 1QP states are $O(1)$ while $W_{\bar{\nu}}^C\approx 0$ for all other levels. In contrast, for $\lambda\approx\lambda_{\rm c}$, one observes a significant transfer from the 1QP to the $n$QP sector for the two levels $|i\rangle$ and $|j\rangle$ so that $W_j\approx 1$ after the anti-level crossing. A large reduced weight in the $n$QP sector is therefore directly linked with the artificial entanglement due to the reduced cluster symmetry. From a certain value of $\ell$, almost all (no) weight is contained in the 1QP ($n$QP) sector which remains true until $\ell_{1n}^{\rm c}$. At this point one observes a significant decrease (increase) of $W_1^C$ ($W_n^C$) and we put $\eta_{1n}=0$ for $\ell\gtrsim\ell_{1n}^{\rm c}$. A large reduced weight in the $n$QP sector is therefore directly linked with the artificial entanglement due to the reduced cluster symmetry.

%|||||||||||||||||||||||||||||||||||||||||||||||||||||||||||||||||||||||||||||||||||||||||||||||||||||||||||||||
%|||||||||||||||||||||||||||||||||||||||||||||||||||||||||||||||||||||||||||||||||||||||||||||||||||||||||||||||

%|||||||||||||||||||||||||||||||||||||||||||||||||||||||||||||||||||||||||||||||||||||||||||||||||||||||||||||||
\begin{figure}
\begin{center}
% NOTICE: lighter JPEG version below. Use EPS one for final version
\includegraphics*[width=\columnwidth]{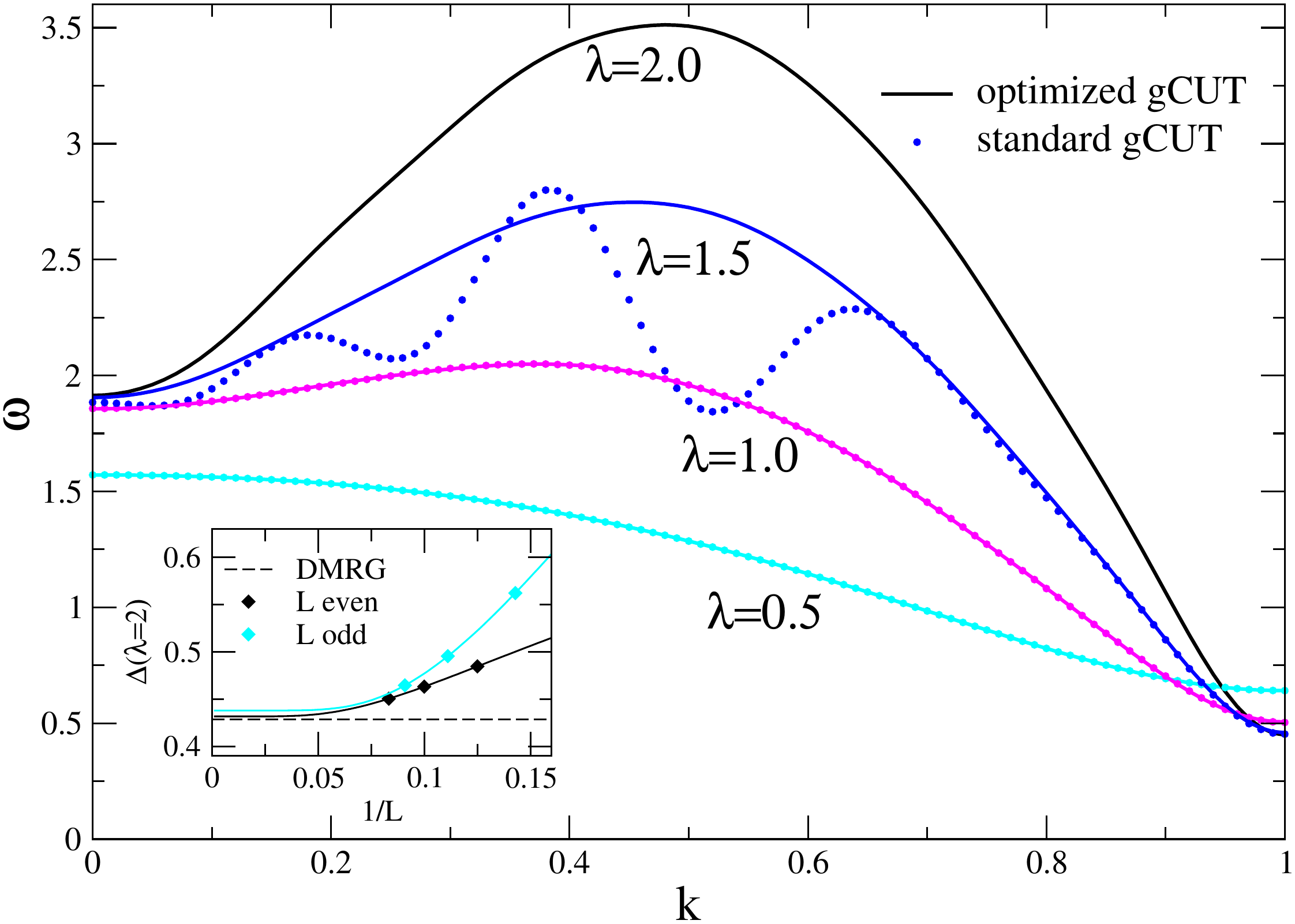}
\end{center}
\caption{{\it (Color online)} One-triplon dispersion $\omega(k)$ of the two-leg Heisenberg ladder for different values of leg exchange $\lambda$ setting the exchange on rungs to $1$. Solid lines (circles) correspond to optimized gCUTs (standard gCUTs) treating graphs up to $L=12$. {\it (Inset)} One-triplon gap $\Delta$ for $\lambda=2$ versus $1/L$. Black (cyan) line is a fit of the form $a_0+a_1\,e^{-a_2\,L}$ using optimized gCUT data with even (odd) $L$. Dashed line represents DMRG data for $\lambda=2$ \cite{Nase14}.} 
\label{fig:2leg_Ladder}
\end{figure}
%|||||||||||||||||||||||||||||||||||||||||||||||||||||||||||||||||||||||||||||||||||||||||||||||||||||||||||||||

%
%
%%%%%%%%%%%%%%%%%%%%%%%
%%%%%%%%%%%%%%%%%%%%%%%
\emph{Applications ---}
%%%%%%%%%%%%%%%%%%%%%%%
%%%%%%%%%%%%%%%%%%%%%%%
%
%
Next we apply the optimized gCUT to two specific quantum spin models. We start with a two-leg antiferromagnetic spin 1/2 Heisenberg ladder, setting the exchange on rungs (legs) of the ladder to $1$ ($\lambda$). This model is gapped for all values of $\lambda$ \cite{Barnes93,Shelton96} and has therefore a finite correlation length $\xi$, but it displays strong interactions among the elementary $S=1$ triplon excitations \cite{Trebst00,Knetter01} and it is therefore an optimal playground to test the performance of the optimized gCUT. 

In practice, we choose to describe the two-leg ladder from the perspective of isolated rungs, i.e.~at $\lambda=0$ one has a product state of singlets as the exact ground state while triplets represent elementary excitations with $S=1$ and energy gap $\Delta=1$. Taking rungs as effective supersites, graphs $\mathcal{G}_L$ of the two-leg ladder corresponds to simple chain segments of length $L$. For each $\mathcal{G}_L$ there are $L$ 1QP reference states with fixed $S^{\rm z}$, since the triplet can be located on each of the $L$ supersites. Here we have treated all $\mathcal{G}_L$ up to $L=12$ extracting the graph-dependent hopping elements of triplon excitations. Embedding the results into the thermodynamic limit and performing a Fourier transformation, yields the one-triplon dispersion $\omega(k)$ displayed in Fig.~\ref{fig:2leg_Ladder} \cite{footnote_even_odd}.

First, we use the standard gCUT scheme with the QP-generator giving suitable results up to $\lambda\approx 1$. This approach fully breaks down for larger values of $\lambda$ which is consistent with the observation that the first artificial anti-level crossing takes place at $\lambda\approx 1.16$ for $L\leq 12$ \cite{footnote_ladder}. This is exactly the situation described above where the one-particle mode is stable in the thermodynamic limit due to momentum conservation. The behaviour is completely different with the optimized gCUT treating the effects of artificial anti-level crossings properly. The results are robust and smooth up to rather large values of $\lambda$, as can be seen in Fig.~\ref{fig:2leg_Ladder}.

Next we turn to the transverse-field Ising model on the square lattice, setting the magnetic field (antiferromagnetic Ising exchange) to $1$ ($\lambda$). In contrast to the two-leg Heisenberg ladder, this model is known to display a zero-temperature quantum phase transition at $\lambda_{\rm crit}=0.3285$ separating the polarized phase at small $\lambda$ from the $\mathbb{Z}_2$ symmetry-broken phase \cite{He90,Oitmaa91,Sachdev11}. We concentrate on the unbroken polarized phase and therefore discuss elementary excitations of this phase in terms of dressed spin flip excitations.

%|||||||||||||||||||||||||||||||||||||||||||||||||||||||||||||||||||||||||||||||||||||||||||||||||||||||||||||||
\begin{figure}
\begin{center}
% NOTICE: lighter JPEG version below. Use EPS one for final version
\includegraphics*[width=\columnwidth]{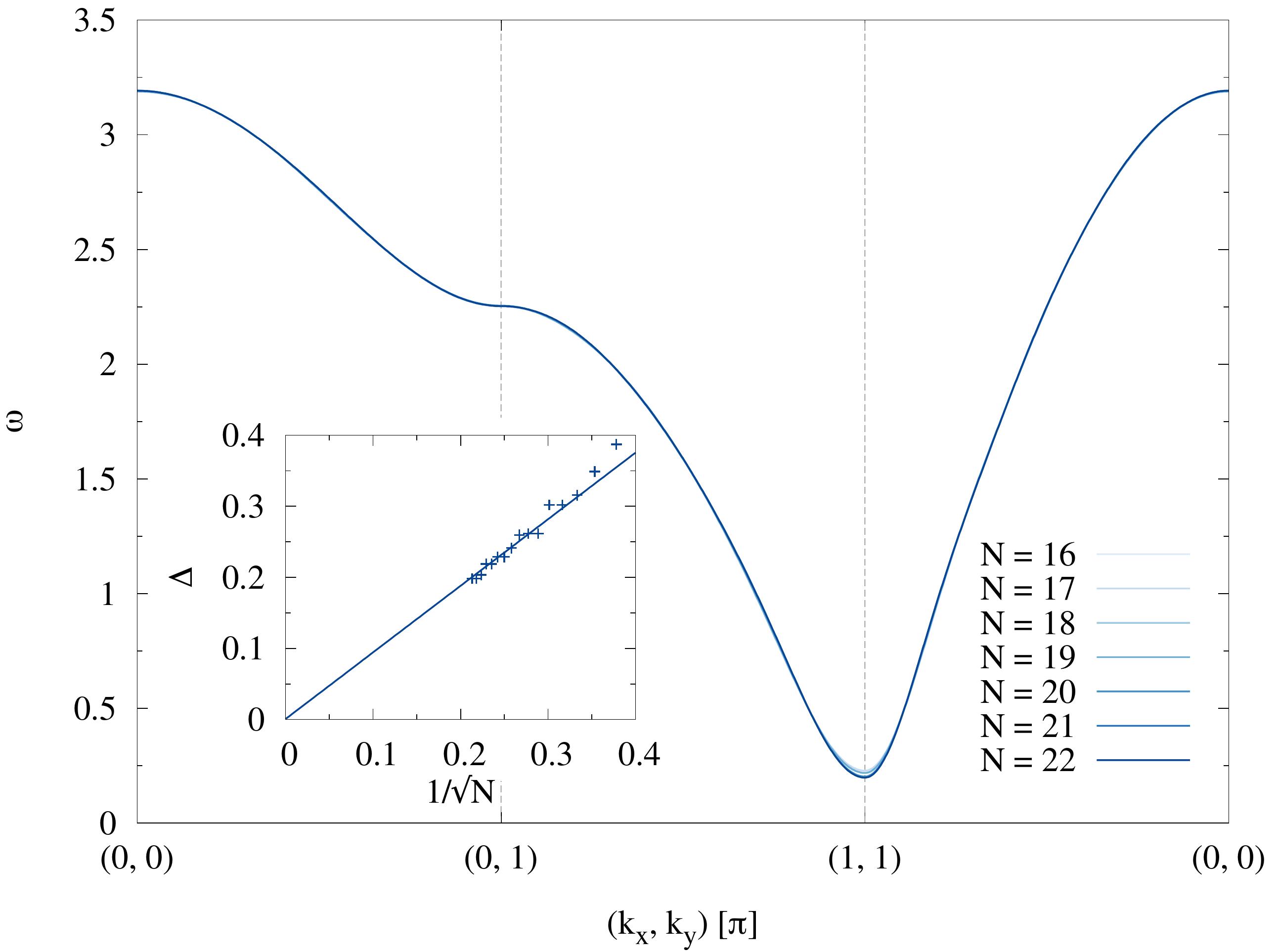}
\end{center}
\caption{{\it (Color online)} One-magnon dispersion $\omega(\vec{k})$ of the transverse-field Ising model on the square lattice at the quantum critical point $\lambda=0.3285$. ({\it Inset}) Scaling of the one-magnon gap $\Delta=\omega(\vec{k}_{\rm af}) $ as a function of $1/\sqrt{N}$    
 reflecting the known exact dynamical exponent $z=1$.} 
\label{fig:TFIM_Square}
\end{figure}
%|||||||||||||||||||||||||||||||||||||||||||||||||||||||||||||||||||||||||||||||||||||||||||||||||||||||||||||||

Here we use a rectangular cluster expansion \cite{Dusuel10,Kallin13} up to $N=22$ spins to calculate the one-particle dispersion $\omega(\vec{k})$ at the quantum critical point $\lambda_{\rm crit}$ as shown in Fig.~\ref{fig:TFIM_Square}, which is by definition the most challenging situation for any NLCE, since $\xi\rightarrow\infty$ and one has gapless excitations at $\vec{k}_{\rm af}=(\pi,\pi)$. Interestingly, we observe a fast convergence with increasing cluster sizes for all momenta being not in the vicinity of $\vec{k}_{\rm af}$. As expected, the gapless nature of the spectrum can only be resolved with a scaling using an appropriate length scale $\mathcal{L}$ of the considered clusters. For the rectangular clusters we take $\mathcal{L}=\sqrt{N}$ \cite{Kallin13} as displayed in the inset of Fig.~\ref{fig:TFIM_Square} for the one-particle gap $\Delta$. Extrapolating in terms of $1/\mathcal{L}$ gives convincing evidence that the gCUT is even capable of catching the correct quantum critical behaviour \cite{footnote_tfim}.

%
%
%%%%%%%%%%%%%%%%%%%%%%%
%%%%%%%%%%%%%%%%%%%%%%%
\emph{Conclusions ---}
%%%%%%%%%%%%%%%%%%%%%%%
%%%%%%%%%%%%%%%%%%%%%%%
%
%
In this work we have identified an inherent complication for NLCEs originating from the reduced graph symmetry leading to \emph{artificial} anti-level crossings when calculating one-particle excitation energies. Our findings are clearly important in a much more general manner, since the same kind of problem is expected to arise for {\it any} separation of degrees of freedom like for many-particle excitations, dynamical correlation functions, or the derivation of effective low-energy models using clusters with reduced symmetry, as done in any NLCE, CORE, or gCUT calculation, e.g.~the derivation of effective spin models in the Mott phase of Hubbard models separating charge and spin degrees of freedom. More generally, effective models are widely used in various fields of quantum physics~\cite{Brandow75,Kvasnicka77,Lindgren82,Durand87}, and their derivation often suffers from the appearance of intruder states in the low-energy spectrum which may cause either discontinuities or spurious behaviours.~\cite{Malrieu1985,Evangelisti1987} Thus, our algorithm might also be used in different fields, for instance quantum chemistry, in order to produce continuous physical results.
%Again, the presence of artificial anti-level crossings leads to entanglement demanding a proper implementation of the generalized notion of cluster additivity as introduced by us. 

Furthermore, it might be worth investigating whether similar problems are also present in high-temperature or non-equilibrium NLCEs as well as in cluster dynamical mean-field theory which all break translational symmetry on clusters. Finally, we are convinced that our optimized gCUT scheme is a potentially useful platform to explore the physics of a vast variety of quantum many-body systems.   

%%%%%%%%%%%%%%%%%%%%%%%%%%%%%%%%%%%%%%%%%%%%%%%%%%%%%%%%%%%%%%%%%%%%%%%%
% Acknowledgement
%%%%%%%%%%%%%%%%%%%%%%%%%%%%%%%%%%%%%%%%%%%%%%%%%%%%%%%%%%%%%%%%%%%%%%%%
\acknowledgments
We acknowledge very useful discussions with Jean-Paul Malrieu and Hong-Yu Yang. This work was in part supported by the Helmholtz Virtual Institute "New states of matter and their excitations".


\begin{thebibliography}{10}
\bibitem{footnote_alpha}
For simplicity, we restrict the discussion to a single type of excitation on supersites $i$, so that no index $\alpha$ appears in the equations.
\bibitem{footnote1}
One may also adapt the generator such that it always follows the upper branch of anti-level crossings for $\lambda\gtrsim\lambda_{\rm c}$, which becomes clearly problematic close to $\lambda_{\rm c}$.
\bibitem{footnote_even_odd}
We observe a strong even-odd effect with respect to $L$ for the two-leg ladder. Consequently, we reduce this effect by taking mean values $(\omega(k)^{(L)}+\omega(k)^{(L-1)})/2$.
\bibitem{footnote_ladder}
Note that results with the standard generator become even worse for $\lambda=2$, so we decided for clarity reasons not to display it explicitly.
\bibitem{footnote_tfim}
For the transverse-field Ising model on the square lattice, no avoided anti-level crossing is detected and the optimized gCUT scheme reduces to the conventional one.
\end{thebibliography}
\end{document}